# Wetting theory for small droplets on textured solid surfaces

Donggyu Kim[1], Nicola M. Pugno[2,3,4], and Seunghwa Ryu[*,1]

**Affiliations**

[1] Department of Mechanical Engineering, Korea Advanced Institute of Science and Technology (KAIST), 291 Daehak-ro, Yuseong-gu, Daejeon 305-701, Republic of Korea

[2] Laboratory of Bio-Inspired and Graphene Nanomechanics, Department of Civil, Environmental, and Mechanical Engineering, University of Trento, Trento, Italy

[3] Center for Materials and Microsystems, Fondazione Bruno Kessler, Trento, Italy

[4] School of Engineering and Materials Science, Queen Mary University of London, Mile End Road, London, United Kingdom

[*] Corresponding author email: ryush@kaist.ac.kr

Conventional wetting theories on rough surfaces with Wenzel, Cassie-Baxter, and Penetrate modes suggest the possibility of tuning the contact angle by adjusting the surface texture. Despite decades of intensive study, there are still many experimental results that are not well understood because conventional wetting theory, which assume an infinite droplet size, has been used to explain measurements of finite-sized droplets. In this study, we suggest a wetting theory that is applicable to any droplet size based on the free energy landscape analysis of various wetting modes of finite-sized droplets on a 2D textured surface. The key finding of our study is that there are many quantized wetting angles with local free energy minima; the implication of this is remarkable. We find that the conventional theories can predict the contact angle at the global free energy minimum if the droplet size is 40 times or larger than the characteristic scale of the surface roughness. Furthermore, we confirm that the pinning origin is the local free energy minima and obtain the energy barriers of pinning as a function of geometric factors. We validate our theory against experimental results on an anisotropic rough surface. In addition, we discuss wetting on a non-uniform rough surface with a rough central region and flat edge. Our findings clarify the extent to which the conventional wetting theory is valid and expand the physical understanding of wetting phenomena of small liquid drops on rough surfaces.

The contact angle is a material property determined by the surface tensions between substrate, liquid, and vapor[1]. Because the materials with extremely small or large contact angles, i.e., with super hydrophilicity or super hydrophobicity, are applicable in many ways, there have been a myriad of studies that have investigated tuning the contact angle via the surface roughness of the substrate based on the conventional wetting theory of rough surface with Wenzel (W), Cassie-Baxter (CB), and Penetrate (P) modes[2-8].

However, the conventional wetting theory[1,9-11] assumes that the liquid droplet is much larger than the characteristic scale of the surface roughness, which frequently is not justifiable in many experiments. First, contact angle predictions using the conventional theory differ from experimental results when the droplet size is small[12-16]. The conventional theory considers a straight boundary between the liquid and the vapor regardless of the liquid tip location and droplet size[1,9-11,17]. However, because the realistic contour of a liquid droplet forms part of a sphere[18], the assumption does not hold when the droplet size to surface texture scale ratio is small. Second, the conventional theory does not account for the pinning mechanism of the liquid tip[14,19-21]. In particular, wetting can involve multiple local free energy minima with different contact angles when pinning occurs, as discussed by Shahraz et al.[22], whereas the conventional theory predicts a single contact angle. Finally, the conventional theory only considers a surface with uniform roughness. There have been several previous studies about the wetting of a surface with non-uniform roughness[17,23], which have asserted that the contact angle is determined by the roughness of the substrate near the liquid tip. However, those studies also either assumed an infinitely large droplet as the conventional theory does or did not explain the phenomena in terms of free energy.

In this work, we present a wetting theory that overcomes the aforementioned limitations of the conventional theory[1,9-11]. The proposed theory is universally valid for any scale of droplets (unless the droplet diameter is a few nanometers or smaller[24]) and reproduces the conventional wetting theory in the limit of an infinite droplet size. To simplify the problem, we consider a finite-sized liquid droplet on a 2D, periodic, rectangular textured surface and compute the free energies of three different wetting modes, W, CB, and P as functions of contact angle $\theta$. The wetting mode at a given contact angle is chosen as the mode with the lowest free energy[25]. We find that the pinning phenomena can be understood by the existence of multiple local minima of the free energy landscape separated by energy barriers. The contact angle at the global minimum recovers the prediction of conventional wetting theory when the droplet size (diameter of the initial droplet) becomes at least 40 times larger than the characteristic scale of the surface roughness (periodicity of the texture). The predictions of our theory is validated against existing experimental results[16]. Due to the existence of many local minima, measured contact angles are not guaranteed to match the prediction of the contact angles at the global free energy minimum but those at local minima. Finally, we calculate the free energy of the wetting on the surface with a non-uniform roughness and reaffirm in terms of free energy that the contact angle is determined by the roughness of the substrate near the liquid tip[17,23].

We adopt and extend the approach of Shahraz et al.[22] and analyse the free energy of a droplet with three wetting modes, W, CB, and P. As illustrated in **Fig. 1a**, we model the rough surface as periodic rectangular textures and model the boundary of the liquid droplet as an arc of a circle. Additionally, to describe the situation where the liquid tip is on the grooves in a mathematically simpler way, we make an approximation that the liquid tip forms a vertical straight line when inside a groove. While such an assumption may cause a small numerical difference, the resulting free energy landscape would correctly captures the multiple local minima and the energy barriers between them qualitatively well. In addition, we find that the free energy has local minima when one end of the droplet is fixed at either the left or right corner of the step. Two cases are illustrated in the inset of **Fig. 1b**. Because a hydrophobic surface prefers to reduce the contact area between the droplet and the substrate at a given contact angle, Case I has the lower free energy for the hydrophobic surfaces ($\theta_e > 90°$), and Case II has the lower free energy for the hydrophilic surfaces ($\theta_e < 90°$) (see Supporting Information for more details). $\theta_e$ refers to the equilibrium contact angle on a flat surface determined by Young's equation[1], $\sigma_{LV} \cos\theta_e = \sigma_{SV} - \sigma_{SL}$, where $\sigma_{LV}, \sigma_{SV}$ and $\sigma_{SL}$ denote the liquid-vapor, solid-vapor and solid-liquid interfacial energies, respectively. For the two cases, the number of grooves below the liquid droplet n can be expressed with the length of the baseline L, width of the step W, and width of the groove G, as described below.

$$n = \begin{cases} \max(n_1, n_1 + \dfrac{\bar{L} - n_1(\bar{G} + \bar{W}) - \bar{W}}{\bar{G}}) & \text{for Case I} \\ \min(n_1 + 1, \dfrac{\bar{L} - n_1(\bar{G} + \bar{W})}{\bar{G}}) & \text{for Case II} \end{cases}$$

where $n_1$ is the number of grooves fully filled with liquid (natural number), and n is defined to account for partially filled grooves (real number). The overbar indicates the dimensionless variables. In what follows, all length and energy variables are normalized against the radius of the circular droplet $R_0$ (i.e. the volume of droplet is $\pi R_0^2$.) and $\sigma_{LV} R_0$, respectively. The values of n are visualized in **Fig. 1b.** for the case when $\bar{G} = 2$ and $\bar{W} = 1$. Thereafter, the radius of the curvature and the free energy of each wetting modes can be derived in terms of n, geometrical factors, and $\theta$, which are presented in Table I. Details on the numerical calculations are described in the Supporting Information.

|    | Curvature Radius $\bar{R}$ | Free Energy $\bar{E}$ |
|----|---|---|
| W  | $\bar{R}_W = \left(\dfrac{\pi - n\bar{G}\bar{H}}{\theta - \sin\theta\cos\theta}\right)^{\frac{1}{2}}.$ | $2\bar{R}_W(\theta - \cos\theta_e \sin\theta) - 2n\bar{H}\cos\theta_e \ (n \in \mathbb{N})$ <br> $2\bar{R}_W(\theta - \cos\theta_e \sin\theta) - 2n_1\bar{H}\cos\theta_e - \bar{H}\cos\theta_e + \bar{H} \ (n \notin \mathbb{N})$ |
| CB | $\bar{R}_C = \left(\dfrac{\pi}{\theta - \sin\theta\cos\theta}\right)^{\frac{1}{2}}.$ | $2\bar{R}_C(\theta - \cos\theta_e \sin\theta) + n\bar{G}(1 + \cos\theta_e).$ |
| P  | $\bar{R}_P = \left(\dfrac{\pi}{\theta - \sin\theta\cos\theta}\right)^{\frac{1}{2}}.$ | $2\bar{R}_P(\theta - \cos\theta_e \sin\theta) + n\bar{G}(-1 + \cos\theta_e).$ |

**Table I. Expressions of curvature radius and free energies of three wetting modes. Overbar indicates dimensionless variables.**

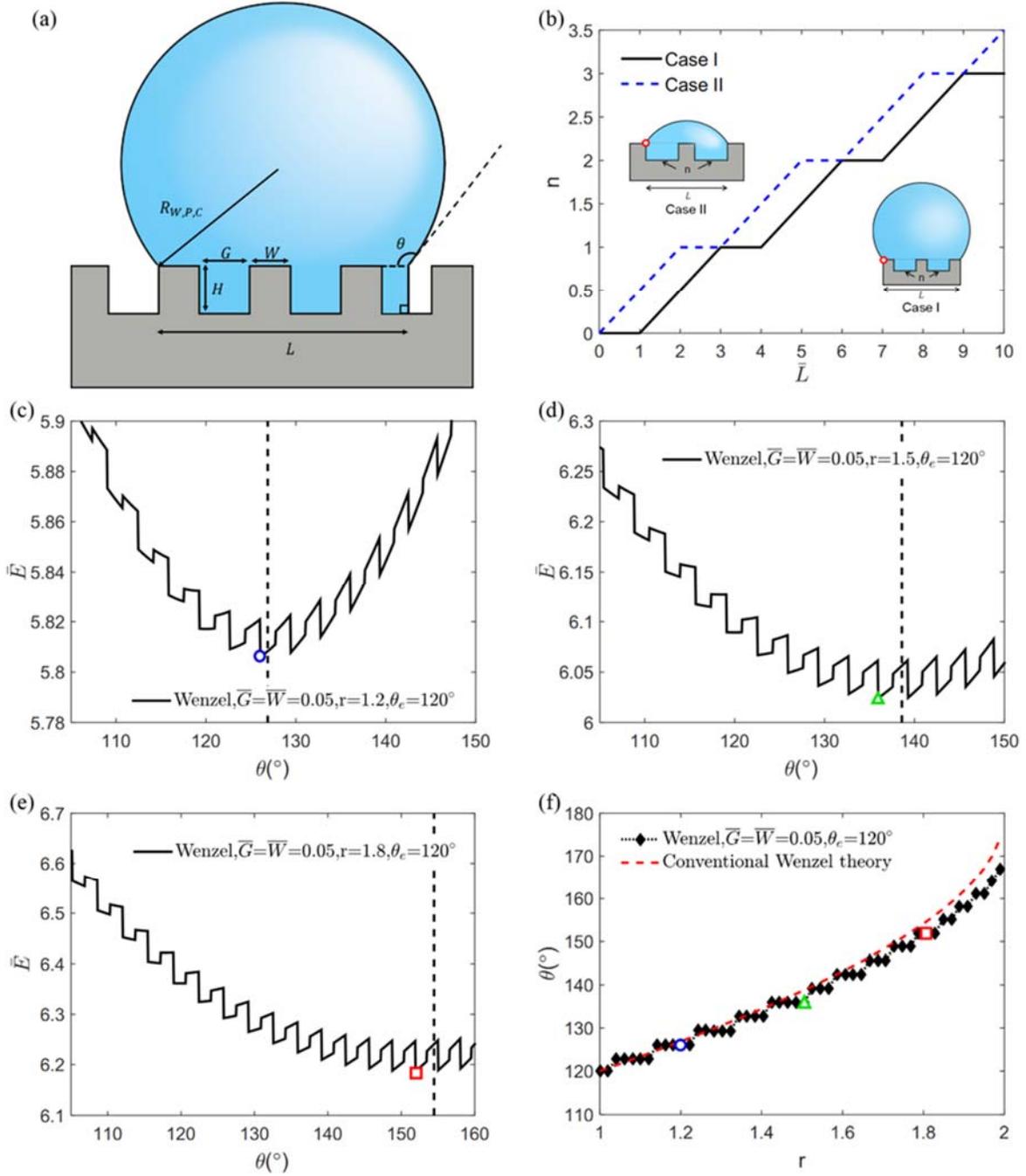

**Figure 1 Assumptions and methods to achieve $r - \theta$ relation.** (a) Schematic of liquid droplet and the 2D textured solid substrate. To simplify the problem, we assume that the liquid droplet has a vertical surface when the non-pinned tip is on the groove (b) Relation between n and $\bar{L}$ when $\bar{G} = 2, \text{and } \bar{W} = 1$ for two locally pinned states. (c-e) Relation between $\theta$ and $\bar{E}$. Global minimum of free energy can be found at various roughness factors r. (f) Relation between r and $\theta$. The contour is constructed by contact angles of the global free energy minima at each roughness factor r.

Based on the free energy expression as a function of $\theta$, we can find the allowable contact angles with free energy minima at a given roughness factor[26] $r = \frac{\bar{G}+\bar{W}+2\bar{H}}{\bar{G}+\bar{W}}$, i.e., the ratio of the true surface area to the projected area. For example, **Figs. 1c~e** show the relationship between the free energy and the contact angle for the W mode when $r = 1.2, 1.5$ and $1.8$, respectively when $\bar{G} = \bar{W} = 0.05$ (i.e., a droplet size of $2R_0$ is 20 times that of the characteristic scale of the surface roughness, $G + W$). The vertical dotted line is the contact angle determined by the conventional wetting theory on the Wenzel mode, $\cos\theta = r\cos\theta_e$. The contact angles at the global free energy minima under different surface roughnesses, which are highlighted by the blue circle, green triangle, and red square, do not match perfectly with the prediction from the conventional wetting theory. Thereafter, one can predict the contact angle as a function of the surface roughness factor $r$ by connecting the contact angles at the global free energy minima, as illustrated in **Fig. 1f**. The conventional wetting theory prediction (red dotted curve), $\cos\theta = r\cos\theta_e$, is also presented in comparison. The contact angles for the CB and P modes can be obtained in a similar way as functions of $f = \frac{\bar{W}}{\bar{G}+\bar{W}}$, which is the fraction of the step area from the projected area.

We then consider how the contact angle changes with the droplet size by varying the dimensionless variable $\bar{G}$, which is the ratio of the groove width to the initial droplet radius. For a given $\bar{G}$, the roughness factor [26] $r$ and the step fraction $f$ can be tuned by changing the height or width of the steps, $\bar{H}$ or $\bar{W}$. The predicted contact angles for W, CB, and P modes are presented in **Figs. 2a-c** with varying values of $r$ and $f$. When $\bar{G} = 0.5$, i.e., the droplet size is 2 times the characteristic scale of the surface roughness, our theory and the conventional theory [9-11] predict different contact angles because the wetting free energy landscape of a small droplet is extremely different from that of a large droplet. Hence, the conventional theory should not be used. However, in case of $\bar{G} = 0.05$, the contact angle from our theory converges to that from the conventional theory within the range of ±10°. When $\bar{G}$ reaches 0.005, the contact angle from our theory recovers the prediction of conventional theory[9-11] almost perfectly, which is expected. Considering the typical resolution of the contact angle measurements (1~2°)[12], we suggest the conventional theory[9-11] should be applied in the case when $\bar{G} \leq 0.025$ (see Supplementary Information for details), i.e., when the droplet size is 40 times bigger than the characteristic scale of the surface roughness.

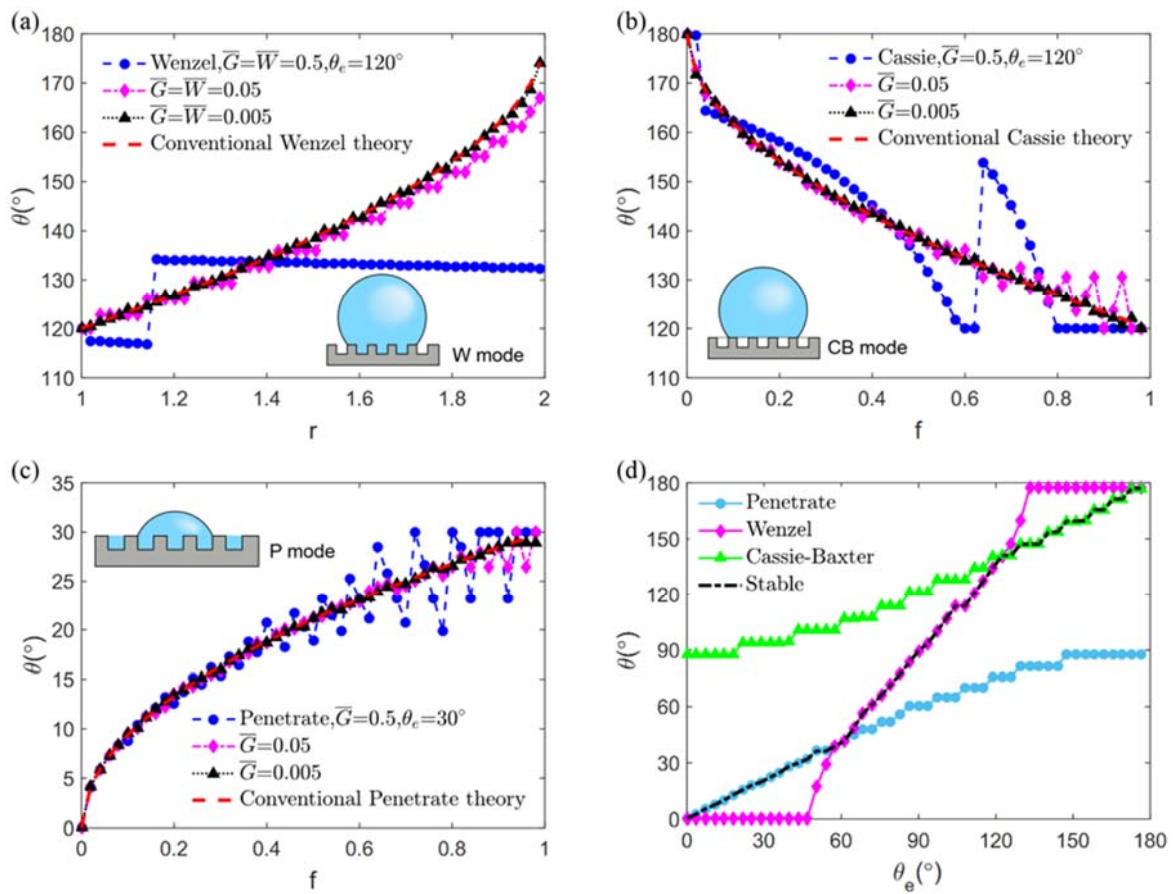

**Figure 2 Results of contact angle and wetting mode prediction from the proposed theory.** (a~c) Convergence curve of each wetting modes( (a): Wenzel, (b): Cassie-Baxter, (c): Penetrate ). Conventional wetting theory should be used when $\bar{G} < 0.025$ (d) Wetting mode selection with respect to $\theta_e$ when $\bar{G} = 0.1$, r = 1.5 and f = 0.5

We can predict the most stable wetting mode for the substrates with different Young's angle $\theta_e$ by comparing the free energies of the three modes[25]. As depicted in **Fig. 3d**, we find that in accordance with the conventional theory, on a hydrophilic surface ($\theta_e < 90°$), the contact angle is given by the higher contact angle between the predictions based on the P mode and W mode and on the hydrophobic surface ($\theta_e < 90°$), by the lower contact angle between the predictions based on the CB mode and W mode. The choice between the W mode and the CB mode is made by comparing the free energy values in **Table I**. To select between the W mode and the P mode, we use the critical contact angle theory[27] instead of the direct free energy comparison because the initial free energy of the P mode differs from the other modes. The critical contact angle, $\theta_C$ is determined when the free energy variation to fill an additional groove is 0. If the contact angle of the flat surface $\theta_e$ is smaller than $\theta_C$, the free energy variation to fill another groove becomes negative, spreading occurs, and the Penetrate mode is selected. In our work, the critical contact angle can be expressed with geometric factors as $\theta_C = \arccos(\frac{\bar{G}}{\bar{G}+2\bar{H}})$. As illustrated in **Fig. 2d**, for the case of $\bar{G} = 0.1, r = 1.5, \text{and } f = 0.5$, one can find that the most stable wetting mode curve follows a similar path as that of the conventional theory.

We then suggest the origin of the pinning effect is the occurrence of multiple local minima in the free energy landscape separated by energy barriers for the three wetting modes. In the W mode, a drastic free energy change occurs when the liquid tip is located near the corner of the step because the additional boundary may form or disappear. The A' and B' points at the end of the step in **Fig. 3a** include an additional boundary line while A and B do not. As depicted in **Fig. 3a**, the free energy differences between A' and A($\Delta\bar{E}_1$) or B' and B($\Delta\bar{E}_2$) act as the primary free energy barrier and form a local free energy minimum. The amount of the energy barrier can be expressed with $\bar{H}$ and $\theta$, as shown in **Table 2**. When the substrate is hydrophobic, the local minimum point is located on B, and one can notice that this corresponds to the experimentally observed liquid tip location[16] when pinning occurs. A similar discussion can be repeated for the Cassie-Baxter or the Penetrate mode. By calculating the free energy of each point of **Figs. 3b and c**, the energy barriers can be formulated, as summarized in **Table 2.** The superscripts A and B refer to the points A and B in the figures, respectively, and the subscripts C and P refer to the wetting modes. Interestingly, while the Wenzel mode always has local free energy minima because of the additional liquid-vapor boundary line, the Cassie-Baxter or the Penetrate mode do not possess local minimum points when $\theta < \theta_e$ or $\theta > \theta_e$, respectively.

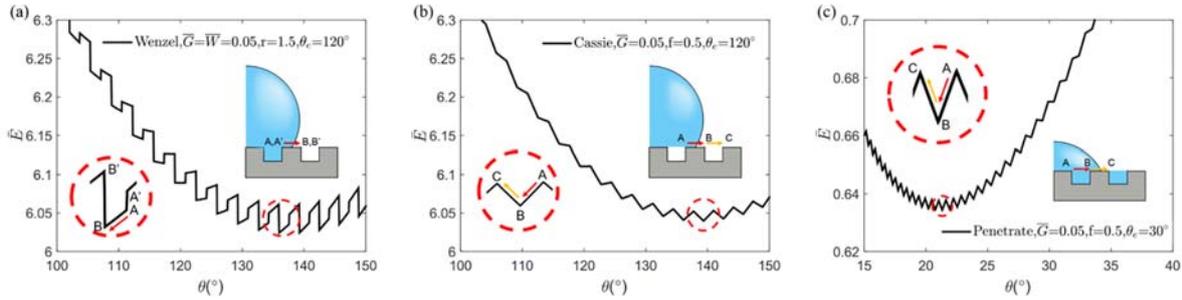

**Figure 3 Relation between $\theta$ and $\bar{E}$ with respect to wetting modes.** (a) Wenzel, (b) Cassie-Baxter, (c) Penetrate. Equations of dimensionless energy barriers are presented in Table 2.

| Wetting Mode | $\Delta\bar{E}_1$ | $\Delta\bar{E}_2$ |
|---|---|---|
| Wenzel | $\bar{H}(1 + \cos\theta_e)$ | $\bar{H}(1 - \cos\theta_e)$ |
| Cassie-Baxter | $2\bar{R}_C^B \theta^B - 2\bar{R}_C^A \theta^A - \bar{W}\cos\theta_e$ | $2\bar{R}_C^B \theta^B - 2\bar{R}_C^C \theta^C - \bar{G}$ |
| Penetrate | $2\bar{R}_P^B \theta^B - 2\bar{R}_P^A \theta^A - \bar{G}$ | $2\bar{R}_P^B \theta^B - 2\bar{R}_P^C \theta^C + \bar{W}\cos\theta_e$ |

**Table 2 Dimensionless energy barriers for each wetting mode**

To validate our theory, we compare our contact angle prediction with existing experimental results[16]. Because our theory is based on a 2D model, we consider the contact angle measurements from the surface with anisotropic roughness. Based on the L, n, $\theta$, and the geometric factors given in the experiments, $R_0$ and $\bar{G}$ can be inferred from the relation among ($\bar{R}_W, R_0, \theta$) for the Wenzel mode (see Supplementary Information for details). Then, we can directly apply our theory to predict the contact angle at the global free energy minimum or local free energy minima for a given $\bar{G}$, as presented in **Table 3.** V, $\theta_{EXP}$, and $n_{EXP}$ refer to the measured values of the liquid volume, contact angle, and the n of the liquid droplet in the experiment, and $\bar{G}$, $\theta_{GM}$, $n_{GM}$, and $\theta_{LM}$ refer to the inferred values at the global minimum (GM) or local minimum (LM), which satisfies $n_{LM} = n_{EXP}$ in the framework of our theory. As illustrated in **Fig. 4b,** the contact angles at the global free energy minimum differ from experimental results by ~12° when $\bar{G} = 0.048$, and this difference decreases as $\bar{G}$ decreases. Interestingly, the measured contact angle coincides with the calculated contact angle at the local free energy minimum where $n_{LM} = n_{EXP}$ is satisfied as depicted in **Fig 4b**. It is interesting to note that even for the $\bar{G} \leq 0.025$ condition, where the contact angle of the global free energy minimum converges to the conventional theory prediction[9-11], the measured contact angles are found to coincide with the local free energy minimum. We find that for a small droplet volume, the difference between the global free energy minimum and a local free energy minimum is relatively small compared with the energy barriers between a pair of local minima, as shown in **Fig. 4a**. As the droplet volume increases, the relative magnitude of the energy barrier decreases, as presented in **Table 2**. This explains why the difference between the measured angle and the contact angle at the global minima decreases as the droplet volume increases (**Fig. 4b**).

| V(mm³) | $\bar{G}$ | $\theta_{EXP}$ | $\theta_{GM}$ | $\theta_{LM}$ | $n_{EXP}$ | $n_{GM}$ | $n_{LM}$ |
|---|---|---|---|---|---|---|---|
| 0.59 | 0.0480 | 140.4 | 151.8 | 140.1 | 14 | 10 | 14 |
| 1.432 | 0.0345 | 143.7 | 154.0 | 143.7 | 18 | 13 | 18 |
| 2.077 | 0.0304 | 144.1 | 153.7 | 144.7 | 20 | 15 | 20 |
| 4.818 | 0.0217 | 148.1 | 154.0 | 148.9 | 25 | 21 | 25 |
| 5.151 | 0.0211 | 149.5 | 153.5 | 149.8 | 25 | 22 | 25 |
| 5.679 | 0.0200 | 150.7 | 153.8 | 151.4 | 25 | 23 | 25 |

**Table 3** Comparison between contacts angles and n for the experimental data, global free energy, and the local free energy. One can notice that the experimental data agree with the local free energy minimum values of our theory.

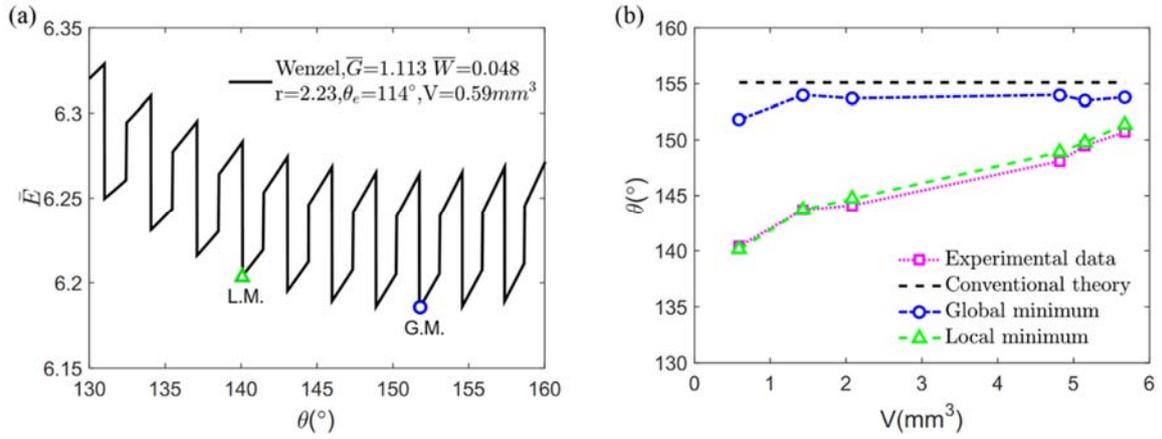

**Figure 4 Comparison between the proposed theory and experimental results.** (a) Relation between $\theta$ and $\bar{E}$ of the Wenzel mode. Experimental data of anisotropic roughness is on the local minimum of the contour (b) Even though the contact angles of the global free energy minimum agree with the conventional theory, the experimental data still contains a gap from the conventional theory because the experimental data is on the local free energy minimum. The gap becomes smaller as the volume of the liquid drop increases.

We then apply our theory to analyse a surface with non-uniform roughness that has a rough center and flat periphery. In other words, we compute the free energy landscape when there exists an upper bound $n_{MAX}$ in the number of filled grooves, $n$. For example, we compute the free energies of the wetting states when $\bar{G} = \bar{W} = 0.05$ and $\theta_e = 120°$ as functions of contact angle $\theta$ with $n_{MAX} = 14$ and 19, as depicted in **Figs. 5a** and **5b**, respectively. **Fig. 5a** illustrates the situation where the area of the rough central region is small enough that the liquid tip is located on the flat region. We find that the global free energy minimum is located at the Young's angle $\theta_e$. It is the case for any $n_{MAX} \leq 14$ because the free energy curve with a fixed $n$ has a minimum at $\theta_e$, as depicted by the red dotted curves in **Fig. 5a**. On the contrary, **Fig. 5b** shows a case where the area of the rough region is large enough and the contact line is located within the rough center. In this case, the local minimum associated with $n = n_{MAX} = 19$ has a higher free energy than the global minimum. Hence, the contact angle prediction becomes identical to the surface with uniform roughness. Our results agrees with the previously proposed wetting theory for a non-uniform rough surface[17,23], which proposed that the contact angle is determined by the roughness condition near the liquid tip. Our work offers an extended theory that enables us to predict whether the tip will be located on the rough center or on the flat region. We note that our theory can be generalized to study the wetting on surfaces with a more complex non-uniform roughness.

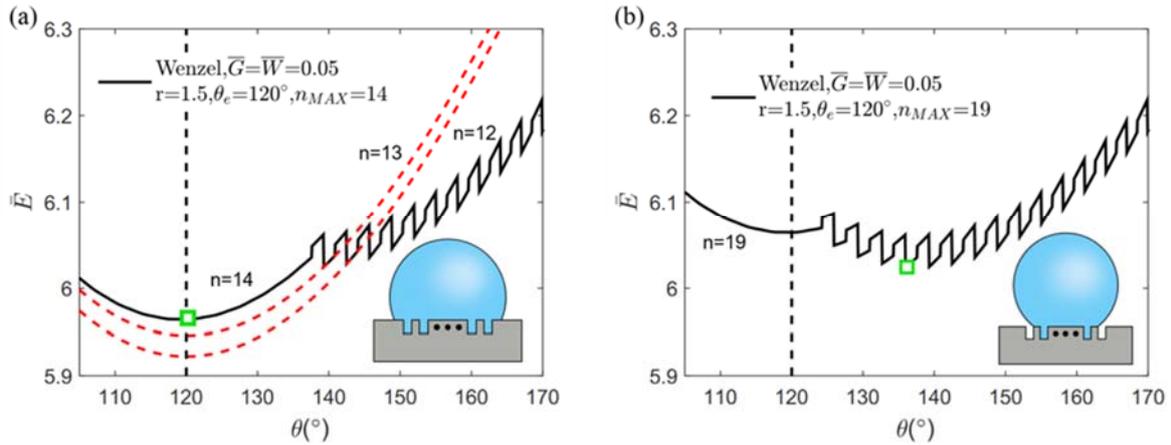

**Figure 5. Relation between θ and $\bar{E}$ of the Wenzel mode when the surface has non-uniform roughness.** The contour contains slices of free energy contours of fixed n (red dotted). The global free energy point is highlighted with the green square symbol (a) The global free energy minimum point might be located on the **n = nMax** curve or (b) the global free energy point of the surface with uniform roughness

In conclusion, we have developed a wetting theory that can predict the wetting angle and wetting mode when the liquid droplet is not much larger than the surface texture scale. Because the conventional theory assumes a much larger size of the droplet compared to the texture scale, there have been limitations in how to analyse experiments that investigate the wetting of small liquid drops. Our theory suggests that conventional theory should be used when the droplet size is at least 40 times larger than the characteristic scale of the surface roughness and provides a deeper physical understanding regarding the wetting of smaller liquid droplets on non-uniform rough surfaces.

**Acknowledgements**

This work is supported by the Basic Science Research Program (2013R1A1A010091) of the National Research Foundation of Korea (NRF) funded by the Ministry of Science, ICT & Future Planning.

**Additional information**

The authors declare no competing financial interests.

**Supplementary Information**

**Wetting theory for small droplets on textured surfaces**

Donggyu Kim[1], Nicola M. Pugno[2], and Seunghwa Ryu[*,1]

**Affiliations**

[1] Department of Mechanical Engineering, Korea Advanced Institute of Science and Technology (KAIST), 291 Daehak-ro, Yuseong-gu, Daejeon 305-701, Republic of Korea

[2] Department of Civil, Environmental, and Mechanical Engineering, University of Trento, Trento, Italy

[*] Corresponding author email : ryush@kaist.ac.kr

**Supplementary Figure 1: Free energy curve of a droplet in Wenzel mode**

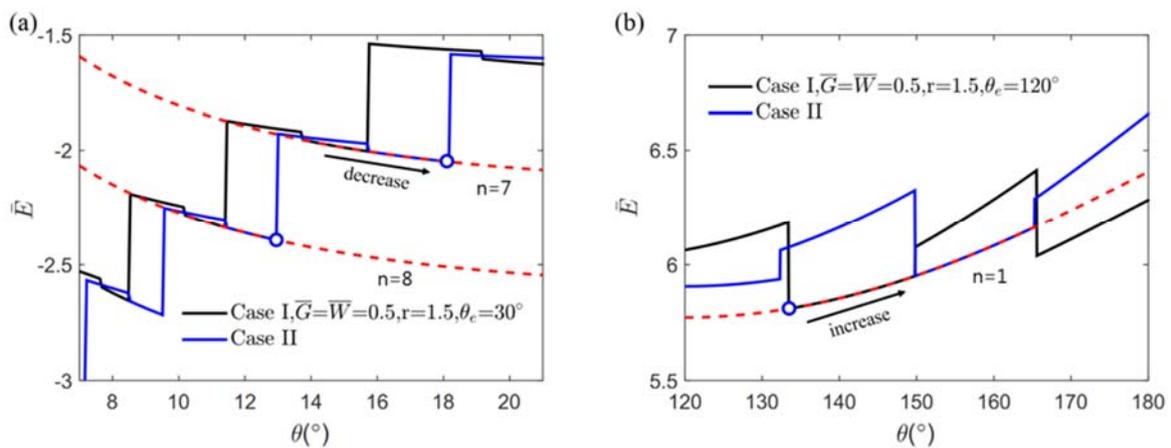

**Supplementary Figure S1**. Convex curvature of Case II is always on the right side of Case I. Because $\bar{E}$ decreases for (a) a hydrophilic substrate and increases for (b) a hydrophobic surfaces as $\theta$ increases, the former has a minimum at Case II and the latter has a minimum at Case I.

## Supplementary Figure 2: Convergence to conventional theory due to $\bar{G}$ for each wetting mode

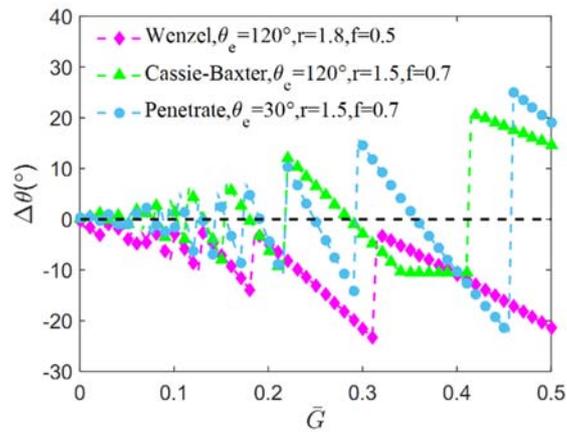

**Supplementary Figure S2**. Convergence to conventional theory due to $\bar{G}$ for each wetting mode. $\bar{W}$ and $\bar{H}$ are set to satisfy the roughness factors (r, f) in the legend. The contact angle difference between the proposed theory and the conventional theory ($\Delta\theta$) converges to $1 \sim 2°$ when $\bar{G} \leq 0.025$

**Supplementary Note 1: The choice between Case I and Case II along wetting modes**

1-1.   Cassie-Baxter & Penetrate mode

From Young's relation[1], one can say the free energy from same area of liquid - solid boundary is always larger than that from the liquid – vapor boundary for any substrate when $\theta > 0°$. Because Case II always contains larger n than in Case I, regardless of L (**Fig. 2b**), Case I contains less area of the liquid-vapor boundary. Consequently, Case I become more stable if the same $\theta$ is assumed. In the case of Penetrate mode, on the contrary, Case II become more stable because the liquid-vapor boundary is substituted by a liquid-liquid boundary, which contains 0 surface energy.

1-2.   Wenzel mode

If the same amount of n is assumed, Case I always has a larger $\bar{L}$ or smaller $\theta$ than Case II. As we noted in the text, the $\bar{E} - \theta$ graph contains convex contours of conserved n, where the minimum is at $\theta_e$ (red dotted in **Figs. S1a**, **S1b**). Because Case I has a larger $\theta$ for conserved n, one can notice the convex contour of Case II is always on the right side of Case I for conserved n. Recalling that the Wenzel mode has a larger equilibrium contact angle than $\theta_e$, when the substrate is hydrophilic, the free energy of the liquid decreases as $\theta$ increases. Therefore, the local free energy minimum near the equilibrium contact angle is on the right side of the contour, Case II. If a hydrophobic substrate is assumed, by the opposite logic, the local free energy minimum is on Case I.

**Supplementary Note 2: Convergence to conventional theory due to $\bar{G}$ for each wetting modes**

As depicted in **Figs. 2a-c**, the predicted contact angle from the proposed theory ($\theta_P$) converges to the contact angle from the conventional theory ($\theta_T$) when $\bar{G} \leq 0.025$. We are not able to find a closed form expression for the difference between the two contact angles ($\Delta\theta = \theta_P - \theta_T$), but find that the envelope of the oscillating $\Delta\theta$ curve decreases with $\bar{G}$ for all wetting modes (**Fig. S2**). The contact angle difference converges with vibration and becomes 1~2° range when $\bar{G} \leq 0.025$.

**Supplementary Note 3: Free energy and curvature radius calculation about θ**

The curvature radius of the liquid droplet of each wetting mode (W,CB,P) can be formulated with a circular trace of the boundary of the liquid and the constant volume constraint as Shahraz et al. [2] reported. In Wenzel mode, the sum of the area of the circular part and of the groove should be constrained. Therefore, the curvature radius of the Wenzel mode $\bar{R}_W$ can be formulated as follows.

$$\bar{R}_W = \left( \frac{\pi - n\bar{G}\bar{H}}{\theta - \sin\theta\cos\theta} \right)^{\frac{1}{2}}$$

After the curvature radius is formulated, the free energy of the droplet can be calculated by summing up the free energy from the liquid-vapor boundary and the liquid-solid boundary as follows.

$$E = \sigma_{LV} A_{LV} + (\sigma_{SL} - \sigma_{SV}) A_{SL}$$

$A_{LV}$ or $A_{SL}$ refer to the area of the boundary between the liquid and the vapor or solid and the liquid. Employing the curvature radius formula, the free energy of the droplet in Wenzel mode can be shown.

$$\bar{E}_W = 2\bar{R}_W(\theta - \cos\theta_e \sin\theta) - 2n\bar{H}\cos\theta_e \quad (n \in \mathbb{N})$$

$$\bar{E}_W = 2\bar{R}_W(\theta - \cos\theta_e \sin\theta) - 2n_1\bar{H}\cos\theta_e - \bar{H}\cos\theta_e + \bar{H} \quad (n \notin \mathbb{N})$$

Because an additional boundary of the liquid-vapor boundary exists when $n \notin \mathbb{N}$ in Wenzel mode, the free energy expression either differs when $n \in \mathbb{N}$ or not. Similar analysis can be repeated for the Cassie-Baxter mode and Penetrate mode to include the curvature radius or free energy of the liquid in CB mode or P mode. In particular, in the case of P mode, we only model the liquid forming the droplet to construct $R_0$ and set the initial state of the free energy with a rough surface in which grooves are filled with liquid for convenience. Then the free energy and the curvature radius of the penetrate mode can be formulated similar with CB mode as follows.

$$\overline{R}_{C,P} = \left(\frac{\pi}{\theta - \sin\theta\cos\theta}\right)^{\frac{1}{2}}$$

$$\overline{E}_{C,P} = \overline{R}_{C,P}(\theta - \cos\theta_e \sin\theta) + n\overline{G}(1 + \cos\theta_e)$$

Because the $n - \theta$ relation is known if the liquid tip condition (Case I or Case II) is suggested, a set of $(\overline{R}, n, \overline{E})$ can be shown with respect to a specific $\theta$. In the case of CB mode and P mode, because $\overline{R}_{C,P}, \overline{E}_{C,P}$, and n are explicitly expressed by $\theta$, a pair of $(\overline{R}_{C,P}, n, \overline{E}_{C,P})$ can be easily obtained for a specific $\theta$. In W mode, $\overline{R}_W, \overline{E}_W$, and n are in an implicit relation; therefore we adopted the bisection method to numerically calculate the set of $(\overline{R}_W, n, \overline{E}_W)$. With this process, the free energy of the droplet can be numerically obtained from a specific $\theta$ and can be used to find the contact angle of the minimum free energy.

**Supplementary Note 4: Comparison with experimental result**

Because the proposed theory is based on a 2D situation, we use an existing experiment[3] that investigate a surface with anisotropic roughness. Assuming the droplet is long enough in the parallel direction to the grooves, an imaginary $R_0$ that preserves the cross-sectional area of the droplet and matches the measured length of the baseline can be inferred from the measured experimental data.

$$R_W^2(\theta_{EXP} - \cos\theta_{EXP}\sin\theta_{EXP}) + n_{EXP}GH = R_0^2\pi \quad \text{(Cross sectional area)}$$

$$L_{EXP} = 2R_W\sin\theta_{EXP} \quad \text{(Baseline length)}$$

The measured value of $\theta_{EXP}, n_{EXP}, G, W, H$ and $L_{EXP}$ for each droplet of different volumes can be found in the reference[3]. With two conditions, $R_0$ and $\bar{G} = \frac{G}{R_0}$ for each droplet can be calculated as in **Table 3**.

After $\bar{G}$ is inferred for each condition, the predicted contact angle from the proposed theory (local minimum and global minimum) can be compared with the experimental result. As mentioned in the text, the free energy curve of the wetting state contains free energy curves with a fixed n having minimum at $\theta = \theta_e$ (**Fig. 5a**). Drawing the free energy curves with different conserved n, the local free energy minimum that satisfies $n_{LM} = n_{EXP}$ or the global free energy minimum point and corresponding $\theta_{LM}$ or $\theta_{GM}$ can be found. As a result, one can observe that the experimental results coincide with $\theta_{LM}$ and the gap between $\theta_{LM}$ and $\theta_{GM}$ decreases as the liquid volume increases.